# Broken electroweak phase at high temperature in the Littlest Higgs model with T-parity

S. Aziz*, B. Ghosh† and G. Dey‡

Department of Physics, University of Burdwan, Burdwan – 713 104, India

**Abstract:** We have examined in detail the nonrestoration of symmetry at high temperature in a finite-temperature Littlest Higgs model, without and with T-parity, by evaluating the one-loop-order finite-temperature integrals of the effective potential numerically, without the high-temperature approximation, $T >> m_i$. We observe that in the model without T-parity it is not possible to find a transition temperature within the allowed temperature range of the model ($0<T<4f$) if the UV completion factors are those which give the Standard Model electroweak minimum, as the effective potential always increases with temperature in the positive direction. However, in the case of the model with T-parity, it is possible to find a transition temperature with the same set of UV completion factors, as, with the increase of temperature, the effective potential decreases in magnitude in the positive side, becomes negative, and increases in magnitude in the negative side, indicating symmetry breaking at high temperature. This type of symmetry nonrestoration at high temperature has been observed earlier in some models involving pseudo Nambu-Goldstone bosons. The behaviour of the *global structure* of the effective potential with T-parity in the theory indicates a strong first order electroweak phase transition, conducive to baryogenesis in the early universe.

*aziz_bu@rediff.com

†ghoshphysics@yahoo.co.in

‡gdey_31@rediffmail.com

## 1. Introduction

The Little Higgs models (LHMs)[1-8] are effective theories at the TeV scale which can describe the Electroweak Symmetry Breaking (EWSB) by stabilizing the Higgs mass against an ultraviolet cut-off by the phenomenon of collective symmetry breaking, which is not possible in the standard model (SM). However, the ultraviolet (UV) completions of the LHMs are not clearly known and therefore conclusions in this model may quite often be uncertain up to reasonable choices or fine tuning of the UV completion factors. Among the LHMs, the Littlest Higgs model($L^2$HM)[9-13] has been found to be an economical theory for the description of the EWSB.

In order to examine other important phenomena associated with the EWSB, *viz.*, Electroweak Phase Transition (EWPT) and baryogenesis, study of finite-temperature effects in LHMs is necessary. In SM, it has been observed that baryogenesis can be explained if there is a strong first-order EWPT[14].

In a recent work [15], based on a modified version[6] of the ($L^2$HM), it has been demonstrated that, contrary to the well-known results of the SM, the electroweak symmetry is not generically restored at high temperature. In the present work, we study in some detail the high temperature behaviour of the finite-temperature effective potential (FTEP) in the $L^2$HM in order to examine the symmetry restoration or non-restoration at high temperature, by numerically evaluating the finite-temperature integrals, as the high-temperature approximation is not valid if the transition temperature is comparable to the masses present in the model. We always work with the un-truncated full non-linear sigma field (NLSF) as that is suitable for studying the global structure of the effective potential for a wide range of the physical Higgs field. Also, in our paper, we extend the earlier works [12,15] by including the features of T-parity in the FTEP.

The paper is organized as follows. In section 2, we review the main features of the $L^2$HM, with and without T-parity. Readers who are familiar with the Littlest

Higgs model can skip this section and go directly to the next section. In section 3, we introduce the full one-loop order effective potential in the L[2]HM, both temperature-independent and temperature-dependent parts. Section 4 contains the detailed expressions of the temperature-independent parts of the effective potentials in various sectors, where we include both the leading quadratic term as well as the logarithmic term. In section 5, we discuss the parameter space of the model taking into account the requirement of the existence of the electroweak minimum at $h = 0.246$ TeV. Section 6 contains detailed analysis of the behaviour of the effective potential against temperatures. Finally, section 7 contains some concluding remarks.

## 2. The Littlest Higgs Model

(a) $L^2$HM without T-parity

The Lagrangian of the Littlest Higgs model is

$$\mathscr{L} = \frac{f^2}{8} Tr(D_\mu \Sigma)(D^\mu \Sigma)^\dagger - \frac{\lambda_1}{2} f \chi_{Li} \varepsilon_{ijk} \varepsilon_{mn} \Sigma_{jm} \Sigma_{kn} u_R - \lambda_2 f \overline{U}_L U_R + h.c. \qquad (2.1)$$

where f is a scale ~ 1 TeV, $\Sigma$ is an NLSF with the covariant derivative,

$$D_\mu \Sigma = \partial_\mu \Sigma - i \sum_{j=1}^{2} [g_j W^a_{j\mu} (Q^a_j \Sigma + \Sigma Q^{aT}_j) + g'_j B_{j\mu} (Y_j \Sigma + \Sigma Y_j)] \qquad (2.2)$$

and

$$\chi_L = \begin{pmatrix} \sigma_2 u_L \\ \sigma_2 b_L \\ U_L \end{pmatrix}, \qquad (2.3)$$

$u_L, b_L, u_R$ being third generation SM quarks and $U_L, U_R$ are new weak-singlet Weyl fermions. In (2.1), $i, j, k$ run among 1,2,3 and $m, n$ run between 4,5. The $L^2$HM has a global SU(5) symmetry which spontaneously breaks to SO(5) giving $\eta, \omega^-, \omega^0, \omega^+, H^-, H^0, H^{0*}, H^+, \phi^{--}, \phi^-, \phi^0, \phi^*_p, \phi^+, \phi^{++}$ as 14 massless Goldstone bosons. The gauging explicitly breaks the SU(5) symmetry at 1 TeV scale by a vacuum condensate which is proportional to

$$\Sigma_0 = \begin{pmatrix} 0 & 0 & 0 & 1 & 0 \\ 0 & 0 & 0 & 0 & 1 \\ 0 & 0 & 1 & 0 & 0 \\ 1 & 0 & 0 & 0 & 0 \\ 0 & 1 & 0 & 0 & 0 \end{pmatrix}. \tag{2.4}$$

And by this symmetry breaking, massive gauge bosons occur by eating up $\eta, \omega^-, \omega^0, \omega^+$. The masses of these gauge bosons are,

$$M_{W_H} = \frac{f}{2}\sqrt{g_1{}^2 + g_2{}^2}\,,\;\; M_{B_H} = \frac{f}{2\sqrt{5}}\sqrt{g_1'^2 + g_2'^2} \tag{2.5}$$

Similarly in the fermion sector we get

$$M_t = 0,\;\; M_T = f\sqrt{\lambda_1{}^2 + \lambda_2{}^2}\,, \tag{2.6}$$

where t is the SM t-quark which has zero mass at the TeV scale, and T is a heavy t-quark. The Higgs boson H is a doublet field,

$$H = \frac{1}{\sqrt{2}}\begin{pmatrix} H^+ \\ H^0 \end{pmatrix} \tag{2.7}$$

Before the electroweak symmetry breaking (EWSB), $H$ is a massless Goldstone boson. After EWSB by the Coleman-Weinberg mechanism[16], $H^0$ gets a VEV and a mass. $H$ can be decomposed as

$$H = \begin{pmatrix} \pi^+ \\ \dfrac{v + h + i\pi^0}{\sqrt{2}} \end{pmatrix} \tag{2.8}$$

where $h$ is the physical Higgs field and $v$ is its VEV. $v=\langle h\rangle=0$ and 246 GeV before and after EWSB respectively. $\pi^+,\pi^-,\pi^0$ are eaten up by the SM gauge bosons which acquire masses at EWSB.

The NLSF has the general structure,

$$\Sigma = e^{2i\Pi/f}\Sigma_0\,, \tag{2.9}$$

where, the 14 Goldstone bosons, mentioned earlier, which are fluctuations around the $\Sigma_0$ field are contained in the $\Pi$ field:

$$\Pi = \begin{pmatrix} -\frac{\omega_0}{2}-\frac{\eta}{\sqrt{20}} & -\frac{\omega^+}{\sqrt{2}} & \frac{H^+}{\sqrt{2}} & -i\phi^{++} & -\frac{i\phi^+}{\sqrt{2}} \\ -\frac{\omega^-}{\sqrt{2}} & \frac{\omega^0}{2}-\frac{\eta}{\sqrt{20}} & \frac{H^0}{\sqrt{2}} & -\frac{i\phi^+}{\sqrt{2}} & \frac{-i\phi^0+\phi_p^*}{\sqrt{2}} \\ \frac{H^-}{\sqrt{2}} & \frac{H^{0^*}}{\sqrt{2}} & \sqrt{\frac{4}{5}}\eta & \frac{H^+}{\sqrt{2}} & \frac{H^0}{\sqrt{2}} \\ i\phi^{--} & \frac{i\phi^-}{\sqrt{2}} & \frac{H^-}{\sqrt{2}} & -\frac{\omega^0}{2}-\frac{\eta}{\sqrt{20}} & -\frac{\omega^-}{\sqrt{2}} \\ \frac{i\phi^-}{\sqrt{2}} & \frac{i\phi^0+\phi_p^0}{\sqrt{2}} & \frac{H^{0^*}}{\sqrt{2}} & -\frac{\omega^+}{\sqrt{2}} & \frac{\omega^0}{2}-\frac{\eta}{\sqrt{20}} \end{pmatrix} \tag{2.10}$$

After EWSB, the $h$ content of $\Pi$ is,

$$\Pi_h = \frac{h}{2}\begin{pmatrix} 0 & 0 & 0 & 0 & 0 \\ 0 & 0 & 1 & 0 & 0 \\ 0 & 1 & 0 & 0 & 1 \\ 0 & 0 & 0 & 0 & 0 \\ 0 & 0 & 1 & 0 & 0 \end{pmatrix} \tag{2.11}$$

and that of $\Sigma$ is

$$\Sigma = e^{2i\Pi_h/f}\Sigma_0 = [1+\frac{i\sqrt{2}}{h}\Pi_h \sin 2\alpha - 4\frac{\Pi_h{}^2}{h^2}\sin^2\alpha]\Sigma_0\,, \qquad (2.12)$$

where, $\alpha = \frac{h}{\sqrt{2}f}$. Writing $s = \sin\alpha$ and $c = \cos\alpha$ we have,

$$\Sigma = \begin{pmatrix} 0 & 0 & 0 & 1 & 0 \\ 0 & -s^2 & i\sqrt{2}sc & 0 & 1-s^2 \\ 0 & i\sqrt{2}sc & 1-2s^2 & 0 & i\sqrt{2}sc \\ 1 & 0 & 0 & 0 & 0 \\ 0 & 1-s^2 & i\sqrt{2}sc & 0 & -s^2 \end{pmatrix} \qquad (2.13)$$

This expression of $\Sigma$ matches with the one given in Ref.11 with changed definitions of the sine and cosine.

(b) $L^2$HM with T-parity (LHT)

Like other little Higgs models, the symmetry structure of $L^2$HM has been enlarged [17] by adding a discrete symmetry called T-parity [18,19], in order to make it consistent with electroweak precision data.

The $\Pi$ field (Eq.2.10) transforms under T-parity as,

$$\Pi \to \Pi' = -\Omega\Pi\Omega\,, \qquad \Omega = diag(1,1,-1,1,1)\,. \qquad (2.14)$$

The effect of this operation in the Higgs sector is to change the sign of the triplet $\phi$ field, keeping that of the doublet H field unchanged. Thus, the T-parity differentiates between the doublet and the triplet Higgs fields. The doublet is T-even and the triplet is T-odd. In a T-parity symmetric theory, these two sectors do not mix up.

In the gauge sector the T-parity operation implies the exchanges, $W_1 \leftrightarrow W_2$, $B_1 \leftrightarrow B_2$ and the invariance of the Lagrangian under these operations demand, $g_1 = g_2$ and $g_1' = g_2'$. With these equalities, the heavy gauge bosons become, $W_H = \frac{1}{\sqrt{2}}(W_2 - W_1)$, $B_H = \frac{1}{\sqrt{2}}(B_2 - B_1)$ which are T-odd and the light gauge bosons become, $W_L = \frac{1}{\sqrt{2}}(W_1 + W_2)$, $B_L = \frac{1}{\sqrt{2}}(B_1 + B_2)$ which are T-even. Thus the T-parity distinguishes between the heavy gauge bosons which get mass at the 1 TeV scale and light gauge bosons or the SM gauge bosons which get mass by EWSB.

In the fermion sector, the Lagrangian with T-parity is of the form,

$$\mathscr{L}_t = \frac{1}{2\sqrt{2}} \lambda_1 f \varepsilon_{ijk} \varepsilon_{xy} [(Q_1{}^{\dagger})_i \Sigma_{jx} \Sigma_{ky} - (Q_2{}^{\dagger} \Sigma_0)_i \tilde{\Sigma}_{jx} \tilde{\Sigma}_{ky}] u_{3R} +$$
$$\lambda_2 f (U_{L_1}{}^{\dagger} \; U_{R_1} + U_{L_2}{}^{\dagger} \; U_{R_2}) + \text{h.c.} \tag{2.15}$$

Where, $\tilde{\Sigma} = \Sigma_0 \Omega \Sigma^{\dagger} \Omega \Sigma_0$ is the image of the $\Sigma$ field under T-parity. $Q_1$ and $Q_2$ are royal SU(3) triplets,

$$Q_1 = \begin{pmatrix} q_1 \\ U_{L_1} \\ 0 \end{pmatrix}, \; Q_2 = \begin{pmatrix} 0 \\ U_{L_2} \\ q_2 \end{pmatrix} \quad \text{with} \quad q_1 = -\sigma_2 \begin{pmatrix} u_{L_1} \\ b_{L_1} \end{pmatrix} \text{ and } q_2 = -\sigma_2 \begin{pmatrix} u_{L_2} \\ b_{L_2} \end{pmatrix}. \tag{2.16}$$

Here the T-parity operation involves interchanges,

$$Q_1 \leftrightarrow -\Sigma_0 Q_2, \quad U_{R_1} \leftrightarrow -U_{R_2} \tag{2.17}$$

The T-parity eigenstates are given by

$$q_{\pm} = \frac{1}{\sqrt{2}}(q_1 \mp q_2) \;,\; U_{L\pm} = \frac{1}{\sqrt{2}}(U_{L_1} \mp U_{L_2}) \,,\; U_{R\mp} = \frac{1}{\sqrt{2}}(U_{R_1} \mp U_{R_2})\,. \tag{2.18}$$

The T-odd states $U_{L-}$ and $U_{R-}$ combine to form a Dirac fermion $T_-$ with $h$-independent mass $M_{T-} = \lambda_2 f$ . The T-odd state $q_-$, having a $h$-independent Dirac mass, is decoupled from the theory [11]. The Lagrangian for the T-even states which take part in collective symmetry breaking is identical to that of the model without T-parity and therefore their contribution to the effective potential will be obtained in the same way as in the case without T-parity. The T-even mass eigenstates are $t_+$, the SM t-quark which get mass after EWSB and heavy top quark which get mass at the 1 TeV scale. These states have $h$-dependent masses.

Another significant consequence of the T-parity implementation in the $L^2$HM is that, the parameter space of the model, consistent with electroweak precision constraints, allows quite small values of the symmetry breaking scale $f$, such as, 400-500 GeV[11].

## 3. The Finite Temperature Effective Potential at One Loop Order

In the imaginary time formalism, the finite temperature effective potential[20] at one loop order has the structure,

$$
\begin{aligned}
V^{(1)} &= \frac{1}{2}\int \frac{d^3p}{(2\pi)^3}\{\sqrt{\vec{p}^2 + M_S{}^2} + 2T\log[1-\exp(-\sqrt{\vec{p}^2 + M_S{}^2}/T)]\} \\
&+ \frac{3}{2}\int \frac{d^3p}{(2\pi)^3}\{\sqrt{\vec{p}^2 + M_V{}^2} + 2T\log[1-\exp(-\sqrt{\vec{p}^2 + M_V{}^2}/T)]\} \\
&- 2\int \frac{d^3p}{(2\pi)^3}\{\sqrt{\vec{p}^2 + M_F{}^2} + 2T\log[1+\exp(-\sqrt{\vec{p}^2 + M_F{}^2}/T)]\}
\end{aligned}
\tag{3.1}
$$

where, S,V and F denote scalar, vector and fermion sectors respectively. In each sector the effective potential separates into a temperature-independent and a temperature-dependent part. The temperature-independent integral as a function of a cut-off momentum $\Lambda$ can be written, up to two leading terms, as

$$
\int \frac{d^3p}{(2\pi)^3}\sqrt{\vec{p}^2 + M^2} = \frac{1}{16\pi^2}[M^2\Lambda^2 + \frac{1}{2}M^4\log\frac{M^2}{\Lambda^2}] \tag{3.2}
$$

In the $L^2$HM, M's can be taken as mass matrices obtained from the Lagrangian (2.1) in arbitrary $\Sigma$ background.

In the gauge sector,

$$
M_V{}^2(\Sigma) = 2f^2\sum_{j=1}^{2}[g_j{}^2(Q_j{}^a\Sigma)(Q_j{}^a\Sigma)^* + g_j'^2(Y_j\Sigma)(Y_j\Sigma)^*] \tag{3.3}
$$

${M_V}^2(\Sigma)$ may be evaluated with $\Sigma_o$ as well as with $\Sigma$. The former gives the contribution to the effective potential at the TeV scale and the latter at the scale of 246 GeV. To get the contribution of mass eigen values we evaluate $TrM_V{}^2(\Sigma)$. Thus we get the gauge boson and fermion contributions to the temperature-independent one-loop order effective potential as,

$$V_G^{(1)}(\Sigma) = \frac{3\Lambda^2}{32\pi^2} TrM_V^2(\Sigma) + \frac{3}{64\pi^2} TrM_V^4(\Sigma)\log\frac{M_V^2(\Sigma)}{\Lambda^2} \tag{3.4}$$

and,

$$V_F^{(1)}(\Sigma) = -\frac{\Lambda^2}{8\pi^2} Tr[M_F{}^\dagger(\Sigma)M_F(\Sigma)]$$
$$-\frac{1}{16\pi^2} Tr[M_F{}^\dagger(\Sigma)M_F(\Sigma)]^2 \log[M_F{}^\dagger(\Sigma)M_F(\Sigma)/\Lambda^2] \tag{3.5}$$

respectively, where,

$$Tr[M_F{}^\dagger(\Sigma)M_F(\Sigma)] = f^2\frac{\lambda_1^2}{2}\varepsilon^{wx}\varepsilon_{yz}\varepsilon^{ijk}\varepsilon_{kmn}\Sigma_{iw}\Sigma_{jx}\Sigma^{*my}\Sigma^{*nz} + h.c. \tag{3.6}$$

Also the scalar fields which get mass contribute to the effective potential. These fields are: $\phi^{++}$, $\phi^{+}$, doubly and singly charged complex scalar fields and $\phi^0$, $\phi_P^0$, scalar and pseudo-scalar neutral fields. Since the Lagrangian has only the gauge and fermion sectors, the masses of the scalar fluctuations have to be calculated within these sectors only, both in the temperature-independent and temperature-dependent parts of the effective potential. That is, the scalar fluctuations will get mass whenever the gauge bosons or fermions will get mass. The mass of $\phi^{++}$, for example, may be calculated by writing,

$$\Pi = \frac{h}{2}\begin{pmatrix} 0 & 0 & 0 & 0 & 0 \\ 0 & 0 & 1 & 0 & 0 \\ 0 & 1 & 0 & 0 & 1 \\ 0 & 0 & 0 & 0 & 0 \\ 0 & 0 & 1 & 0 & 0 \end{pmatrix} + \begin{pmatrix} 0 & 0 & 0 & -i\phi^{++} & 0 \\ 0 & 0 & 0 & 0 & 0 \\ 0 & 0 & 0 & 0 & 0 \\ i\phi^{--} & 0 & 0 & 0 & 0 \\ 0 & 0 & 0 & 0 & 0 \end{pmatrix} = \Pi_1 + \Pi_2 ,$$

$$\Sigma = e^{2i\Pi/f}\Sigma_0 = e^{2i/f(\Pi_1+\Pi_2)}\Sigma_0 = e^{2i\Pi_1/f}e^{2i\Pi_2/f}\Sigma_0 = \Sigma_h(1+\frac{2i}{f}\Pi_2 - \frac{2}{f^2}\Pi_2{}^2)\Sigma_0 ,$$

where, $$\Sigma_h = \begin{pmatrix} 1 & 0 & 0 & 0 & 0 \\ 0 & 1-s^2 & i\sqrt{2}sc & 0 & -s^2 \\ 0 & i\sqrt{2}sc & 1-2s^2 & 0 & i\sqrt{2}sc \\ 0 & 0 & 0 & 1 & 0 \\ 0 & -s^2 & i\sqrt{2}sc & 0 & 1-s^2 \end{pmatrix} \qquad (3.7)$$

and evaluating the coefficient of $\Pi_2{}^2$ in the gauge boson and fermion sectors. In our formulation, all the scalar fields mentioned above have the same contribution to the effective potential.

In the present work, the finite temperature integrals in (3.1) are evaluated numerically. The expressions of the temperature-independent parts of the effective potentials are given in the following section.

## 4. Expressions of the Temperature–Independent Parts of the Effective Potentials.

In the gauge boson sector, the real and finite part of the temperature-independent effective potential, obtained from (3.4), is

$$V_G^{(1)}(\Sigma) = af^4[\{\tfrac{9}{4}(g_1{}^2 + g_2{}^2) + \tfrac{3}{20}(g_1'^2 + g_2'^2) + (g_1{}^2 + g_2{}^2 + g_1'^2 + g_2'^2)s^4\}$$
$$+(3/1024\pi^2)\{g_1{}^4\log(g_1{}^2/64\pi^2) + g_2{}^4\log(g_2{}^2/64\pi^2)\}(s^2-4)$$
$$+(3/640000\pi^2)\{5(g_1'^4\log(3g_1'^4/51200\pi^4) + g_2'^4\log(3g_2'^4/51200\pi^4))s^2 +$$
$$4(g_1'^4\log(19g_1'^2/6400\pi^2) + g_2'^4\log(19g_2'^2/6400\pi^2)(40s^2+1) + 2(g_1'^4\log(g_1'^2/400\pi^2) +$$
$$g_2'^2\log(g_2'^2/400\pi^2)(5s^2+8)\}]$$

(4.1)

where, the first line is from the leading quadratic term and the rest from the logarithmic term. In the latter we have taken up to the leading order in s both in the SU(2) and U(1) sectors.

In the fermion sector we have,

$$V_F^{(1)}(\Sigma) = a'f^4[\{-48\lambda_1^{\ 2}/(4\lambda_1^{\ 2}-\lambda_t^{\ 2})+24\lambda_1^{\ 2}s^4\}$$
$$-\{2(\sqrt{2}\lambda_1^{\ 2}-\lambda_t\lambda_1)/\sqrt{4\lambda_1^{\ 2}-\lambda_t^{\ 2}}\}^4\log\{(\sqrt{2}\lambda_1^{\ 2}-\lambda_t\lambda_1)/4\pi\sqrt{4\lambda_1^{\ 2}-\lambda_t^{\ 2}}\}$$
$$-\{\lambda_1^{\ 2}(6\lambda_1^{\ 2}-\lambda_t^{\ 2}+4s\sqrt{4\lambda_1^{\ 2}-\lambda_t^{\ 2}})/(4\lambda_1^{\ 2}-\lambda_t^{\ 2})\}^2$$
$$\times\log\{\lambda_1^{\ 2}(6\lambda_1^{\ 2}-\lambda_t^{\ 2}+4s\sqrt{4\lambda_1^{\ 2}-\lambda_t^{\ 2}})/16\pi^2(4\lambda_1^{\ 2}-\lambda_t^{\ 2})\}]$$

(4.2)

where,

$$\lambda_t = \frac{\sqrt{2}\lambda_1\lambda_2}{\sqrt{\lambda_1^{\ 2}+\lambda_2^{\ 2}}} \tag{4.3}$$

is the SM top quark Yukawa coupling constant. Again in the logarithmic term we have written up to the leading order term in s.

For the four scalar fields, $\phi^{++},\phi^{+},\phi^{0},\phi_P^{0}$ together we have

$$V_S^{(1)}(\Sigma) = af^4[2(g_1^{\ 2}+g_2^{\ 2}+g_1'^2+g_2'^2)(1-4s^4)$$
$$+(27/32\pi^2)(g_1^{\ 2}+g_2^{\ 2})^2\log(3(g_1^{\ 2}+g_2^{\ 2})/32\pi^2)(2-s^4)$$
$$+(3(g_1'^2+g_2'^2)^2/40000\pi^2)((40s^2+16)\log((g_1'^2+g_2'^2)/200\pi^2)$$
$$+(260s^2-961)\log(31(g_1'^2+g_2'^2)/800\pi^2)$$
$$+s^2\{930\log((g_1'^2+g_2'^2)/80\pi^2)-620\log(3(g_1'^2+g_2'^2)/800\pi^2)\}]$$
$$+a'f^4[128\lambda_1^{\ 2}(1-4s^4)-4\lambda_1^{\ 4}s^2\{1535\log(0.81\lambda_1^{\ 2})\}/\pi^2]$$

(4.4)

In the equations (4.1), (4.2) and (4.4) $a$ and $a'$ are the UV completion factors in the gauge and fermion sectors respectively, whose values should be of the order of unity. The UV factors are necessary due to the use of the upper cut off in the loop momentum integrals. The total temperature-independent potential is,

$$V^{(1)}(\Sigma) = V_G{}^{(1)}(\Sigma) + V_F{}^{(1)}(\Sigma) + V_S{}^{(1)}(\Sigma) \tag{4.5}$$

## 5. Parameter Space

Without specification of the UV Physics above the cut off the Higgs potential and the couplings are determined at the one loop order by nine parameters of the effective theory below the cut off /2/. These are $g_1, g_2, g_1', g_2', \lambda_1, \lambda_2, f, a$ and $a'$. $g_1, g_2, g_1', g_2'$ are related to the SM weak and hypercharge gauge couplings as,

$$g = \frac{g_1 g_2}{\sqrt{g_1^2 + g_2^2}} \ , \quad g' = \frac{g_1' g_2'}{\sqrt{g_1'^2 + g_2'^2}} \tag{5.1}$$

Defining,

$$G^2 = g_1{}^2 + g_2{}^2 \text{ and } G'^2 = g_1'^2 + g_2'{}^2, \tag{5.2}$$

we get,

$$g_1^2, g_2^2 = \frac{G^2}{2} \pm \sqrt{G^2 - 4g^2} \ , \ g_1'^2, g_2'^2 = \frac{G'^2}{2} \pm \sqrt{G'^2 - 4g'^2} \tag{5.3}$$

Since $g$ and $g'$ are known from SM as, $g = \sqrt{4\pi\alpha} / \sin\theta_W$, $g' = g \tan\theta_W$, $g_1, g_2, g_1', g_2'$ can be fixed if $G$ and $G'$ are taken as free parameters of the model along with the constraints (5.2). Also the reality conditions from (5.3) show that,

$$G \geq 2g \text{ and } G' \geq 2g'. \tag{5.4}$$

In the top sector, $\lambda_1$ and $\lambda_2$ are related to the SM Yukawa coupling $\lambda_t$ by the equation (4.3). If $\lambda_1$ is taken as a parameter of the model, then $\lambda_2$ can be determined from the relation,

$$\lambda_2 = \frac{\lambda_1 \lambda_t}{\sqrt{4\lambda_1^2 - \lambda_t^{\ 2}}} . \qquad (5.5)$$

The reality condition of $\lambda_2$ in (5.5) shows that,

$$\lambda_1 \geq \lambda_t / 2 \qquad (5.6)$$

Thus with the help of the SM parameters, the number of parameters in L[2]HM can be reduced as, $g_1, g_2, g'_1, g'_2, \lambda_1, \lambda_2 \;\rightarrow\; G, G', \lambda_1$. Then, taking plausible values of $f$, the values of $a$ and $a'$ can be obtained in a definitive way from the requirement of getting the electroweak minimum at $h = 246\,\text{GeV}$. This procedure is inspired by the fact that the LHMs are minimal extensions of the SM[8,9]. The method of getting the SM electroweak minimum in the allowed parameter space is described below.

Using the SM value, $g = 0.63$ and $g' = 0.344$ we get from (5.4),

$$G \geq 1.26\ ,\ \ G' \geq 0.69 . \qquad (5.7)$$

Also, since $\lambda_t = 1.02$ we get from (5.6),

$$\lambda_1 \geq 0.51 \qquad (5.8)$$

Taking, $G = 1.3$, $G' = 0.7$, $\lambda_1 = 0.55$, $g_1^{\ 2} = \frac{G^2}{2} + \sqrt{G^2 - 4g^2}$ =1.165, $g_2^{\ 2} = G^2 - g_1^{\ 2} = 0.525$, $g_1'^2 = \frac{G'^2}{2} + \sqrt{G'^2 - 4g'^2}$ =0.374, $g_2'^2 = G'^2 - g_1'^2 = 0.116$, $\lambda_1 = 0.55$ and $f = 1$, we get from the condition,

$$\left( \frac{\partial V^{(1)}(\Sigma)}{\partial s} \right)_{s=0.173} = 0 \qquad (5.9)$$

$a' = 0.004a$. Further derivative of $V^{(1)}(\Sigma)$ gives

$$\left(\frac{\partial^2 V^{(1)}(\Sigma)}{\partial s^2}\right)_{s=0.173} = -2.206a\,, \tag{5.10}$$

where, s=0.173 corresponds to the value of $h$, at which the electroweak minimum occurs. As a minimum corresponds to a positive value of second derivative, to get the electroweak minimum we can make the choice, $a = -1,\ a' = -0.004$. However, we observe that by making this choice we can not keep both $a$ and $a'$ to be of the order of unity. This is consistent with an earlier finding [12] that realistic values of $f$ and $m_h$ ( the higgs mass) forces the UV parameters to be different by two orders of magnitude.

It may be noted that nowhere in our analysis we are taking into consideration the triplet higgs field because its VEV is less than 0.015 TeV for $f = 1$ TeV [10] and it is unlikely to influence the global structure of the effective potential in the $h$ direction *i.e.*, $t = 0$ direction, which we are actually interested in.

In the case of LHT, since $g_1 = g_2$ and $g_1' = g_2'$, corresponding to the minimum values of $G$ and $G'$ we have, $g_1 = g_2 = g/\sqrt{2}$ and $g_1' = g_2' = g'/\sqrt{2}$ respectively. Using these values of $g_1, g_2, g_1', g_2'$ and $f = 0.5$ TeV and following the above procedure of determining UV completion factors, we get a possible set of values, $a = -1,\ a' = -0.02.$ It is to be noted that a reduction in the differences in the magnitudes of $a$ and $a'$ results from a smaller value of $f$, which will prevent the higgs mass $m_h$ to be unnaturally much larger than its current lower bound ( ~115 GeV ), since $m_h$ is $O(f)$. It is also to be noted that implementation of T-parity and consequent smaller value of $f$ (~0.5 TeV) makes the LHMs completely natural [19,20]

## 6. Results and Discussions

We have examined the properties of the one-loop order effective potential in the $L^2$HM, both without and with T-parity. The finite-temperature integrals have been evaluated using the MATHEMATICA software version 6.0.

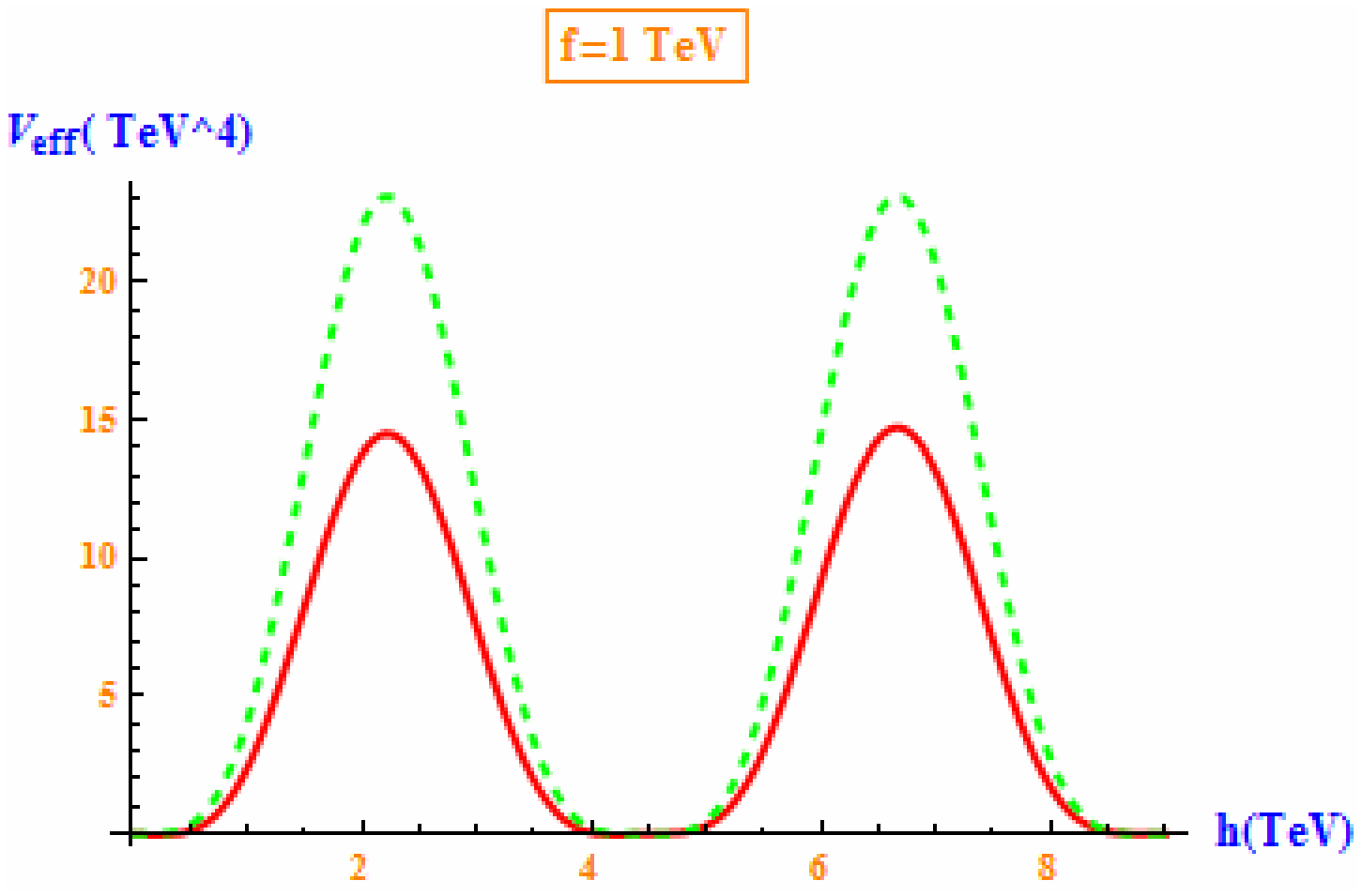

**Fig.1 The global structures of the zero-temperature part of the tree-level (green, dotted) and one loop order (red, solid) effective potential** $V_{eff}$=V(h)-V(h=0) **with our chosen parameter set.**

In Fig.1 we have shown the global structure of the temperature-independent part of the effective potential both at the tree-level[15] and the one-loop level (Eq. (4.5)) for $f = 1$TeV, calculated with the $\Sigma$ given in Eq.(2.13). The periodicity in the global structure is the result of the periodicity in $\Sigma$ in Eq.(2.12) in terms of which, the mass matrices and the effective potentials are written. The tree-level potential has been obtained with the values of the UV completion factors, $a = 1$, $a' = 1$. As has been discussed in the previous section, in the case of one-loop order effective potential, the values of the UV completion factors are so chosen as to give the SM electroweak minimum at h=0.246 TeV at zero temperature. Although, the values obtained in this way, *viz.*, $a = -1$, $a' = -0.004$ go against the spirit of the $L^2$HM, in so far as the small value of $a'$ is concerned, it might indicate some significant modification in the fermion sector of the model, once the UV physics is clearly known. As has been pointed in Ref. 12, values of $f$ and $m_h$ in the desired range is obtained only with values of the UV completion factors which differ by two orders of magnitude.

In Fig.2, we show more clearly the tree and one-loop order effective potential near the SM electroweak minimum.

In Fig.3, we examine the temperature variation of the one-loop order effective potential near the SM electroweak minimum, starting from zero temperature to the maximum possible value allowed by the model, *viz.*, $T = 4f$, keeping the values of $a$ and $a'$ fixed at those values which give the SM electroweak minimum at zero temperature. Here, we have numerically evaluated the finite temperature integrals without the high-temperature approximation. Clearly, an increase in temperature in the finite-temperature $L^2HM$ can not restore the SM electroweak symmetry, unless the UV completion factors also change with temperature and become positive at some point. A finite-temperature version of the UV completion theory may give credence to this issue. However, our result is consistent with that obtained with the modified version of $L^2HM$ in Ref.15, where a maximum at $h=0$ in the allowed temperature range has been obtained. The usual SM symmetry restoration at much lower temperature (T~100 GeV) may take place after the heavy particles in the model get decoupled. However, one should also take into consideration the fact that in the $L^2$HM, the zero-temperature one-loop order effective potential has a quadratic divergence whereas in the SM there is a logarithmic divergence in such case.

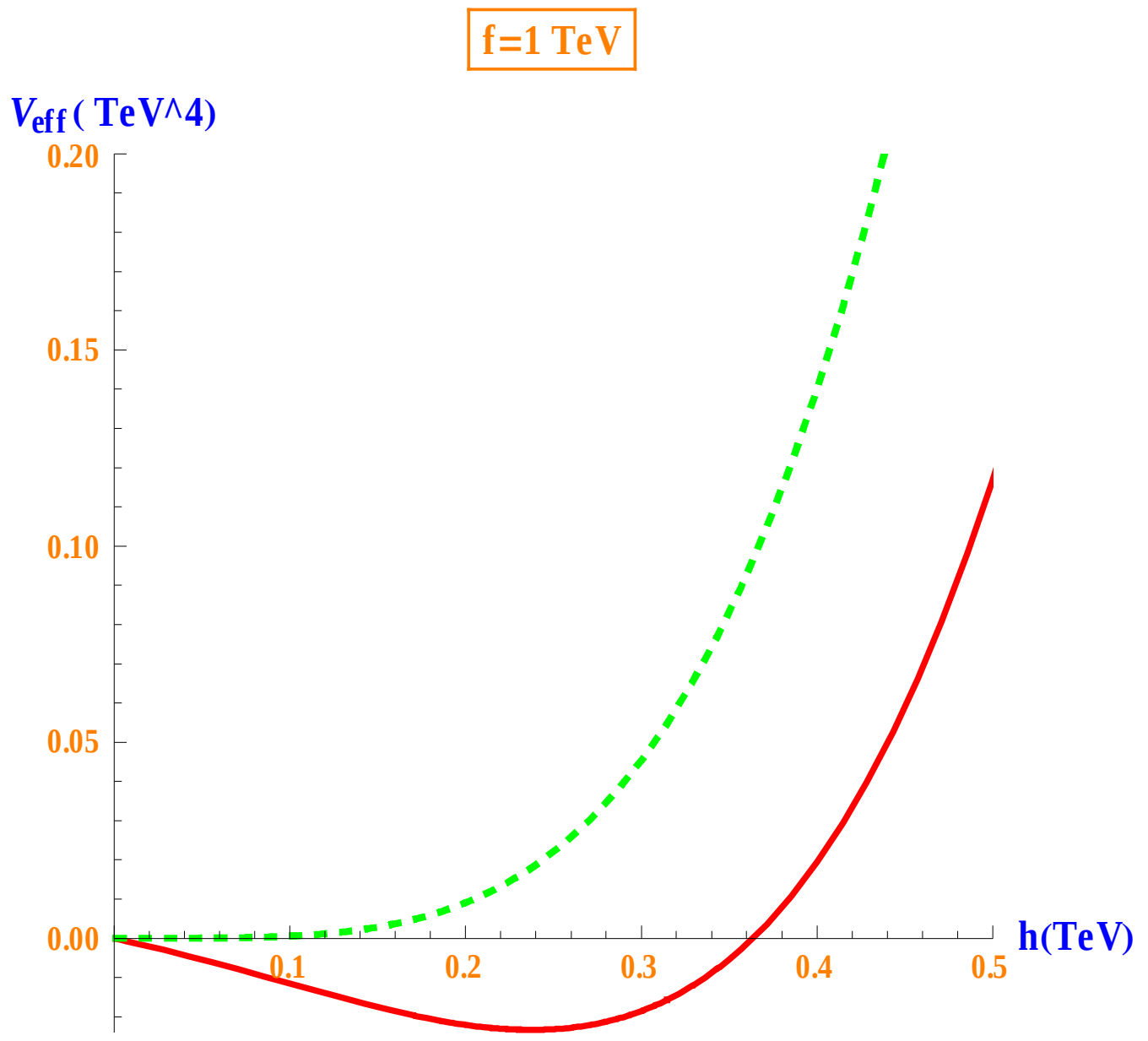

**Fig. 2 The zero-temperature tree-level (green, dotted) and one loop order (red, solid) effective potentials near the SM electroweak minimum.**

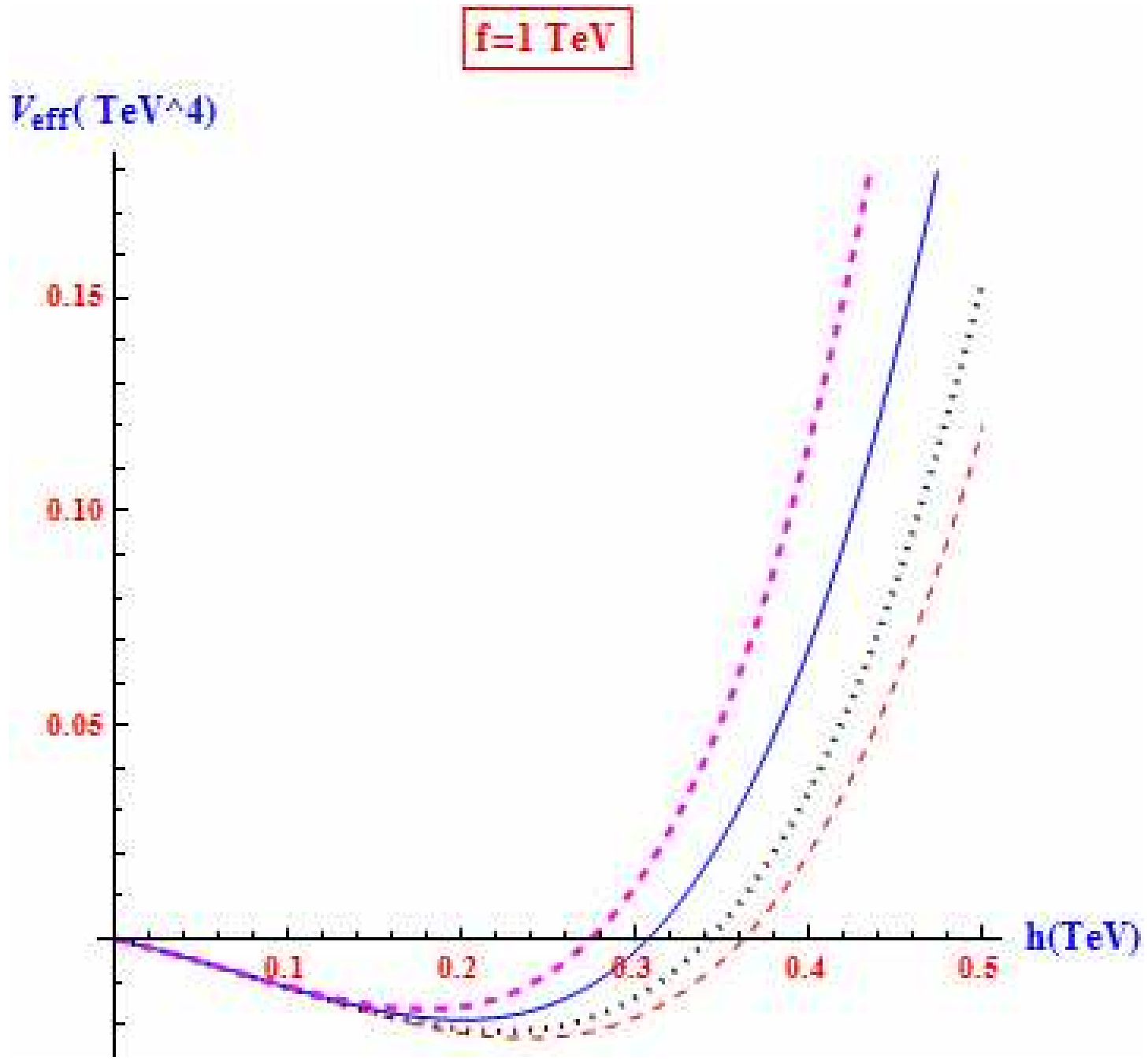

**Fig.3 Finite-temperature one loop order effective potential near the SM electroweak minimum at T=0 TeV(red, dashed), T=2 TeV(black, dotted), T=3 TeV(blue, solid) and T=4 TeV(magenta, thick-dashed). Here the UV completion factors are fixed at their values at the zero-temperature.**

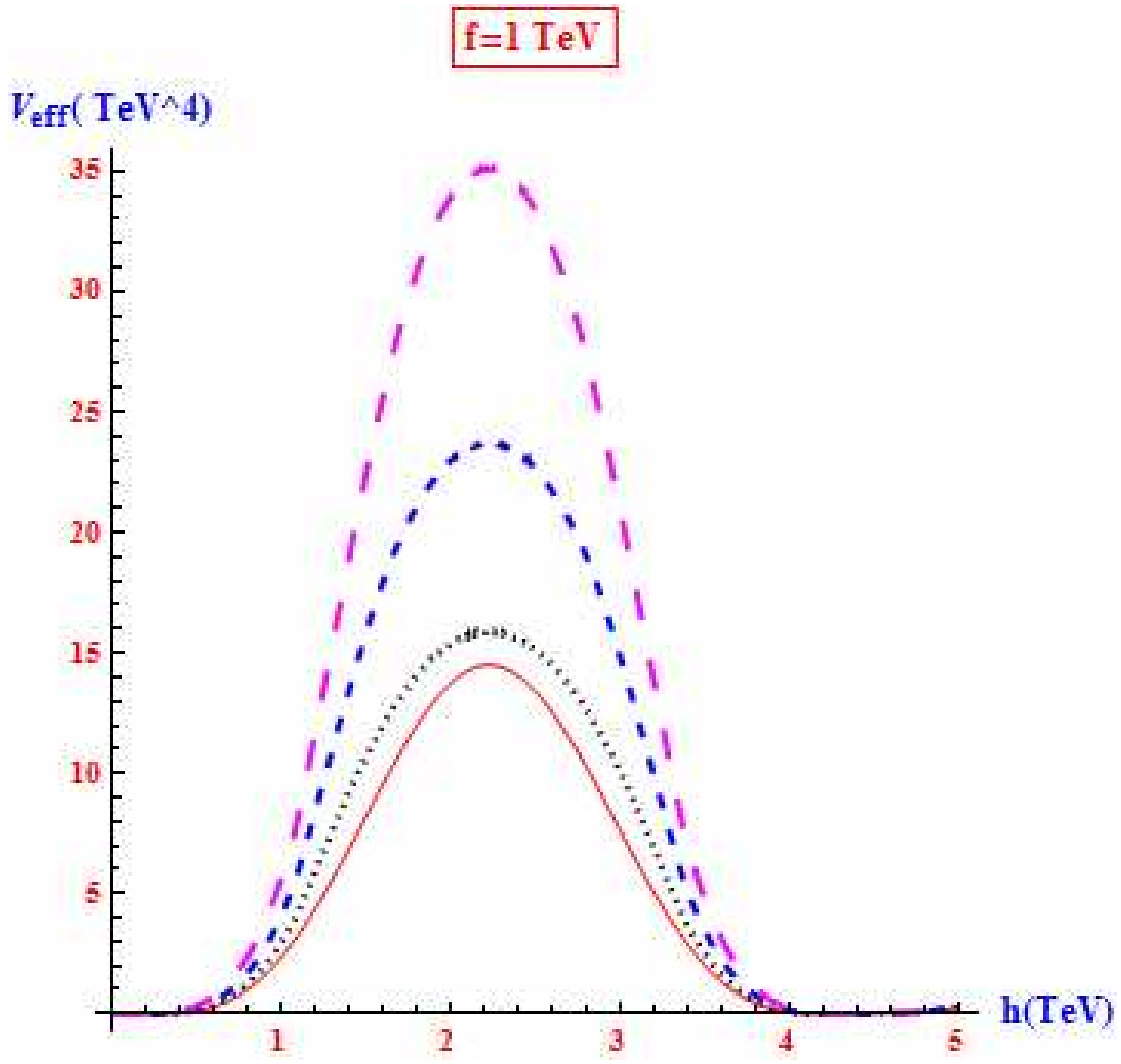

**Fig.4 Global structure of the finite-temperature one loop order effective potential (from bottom to top) at T=0 TeV (red, solid), T=2 TeV(black, dotted), T=3 TeV(blue, medium-dashed) and T=4 TeV(magenta, large-dashed). Here the UV completion factors are fixed at their values at the zero-temperature, chosen to obtain the SM electroweak minimum.**

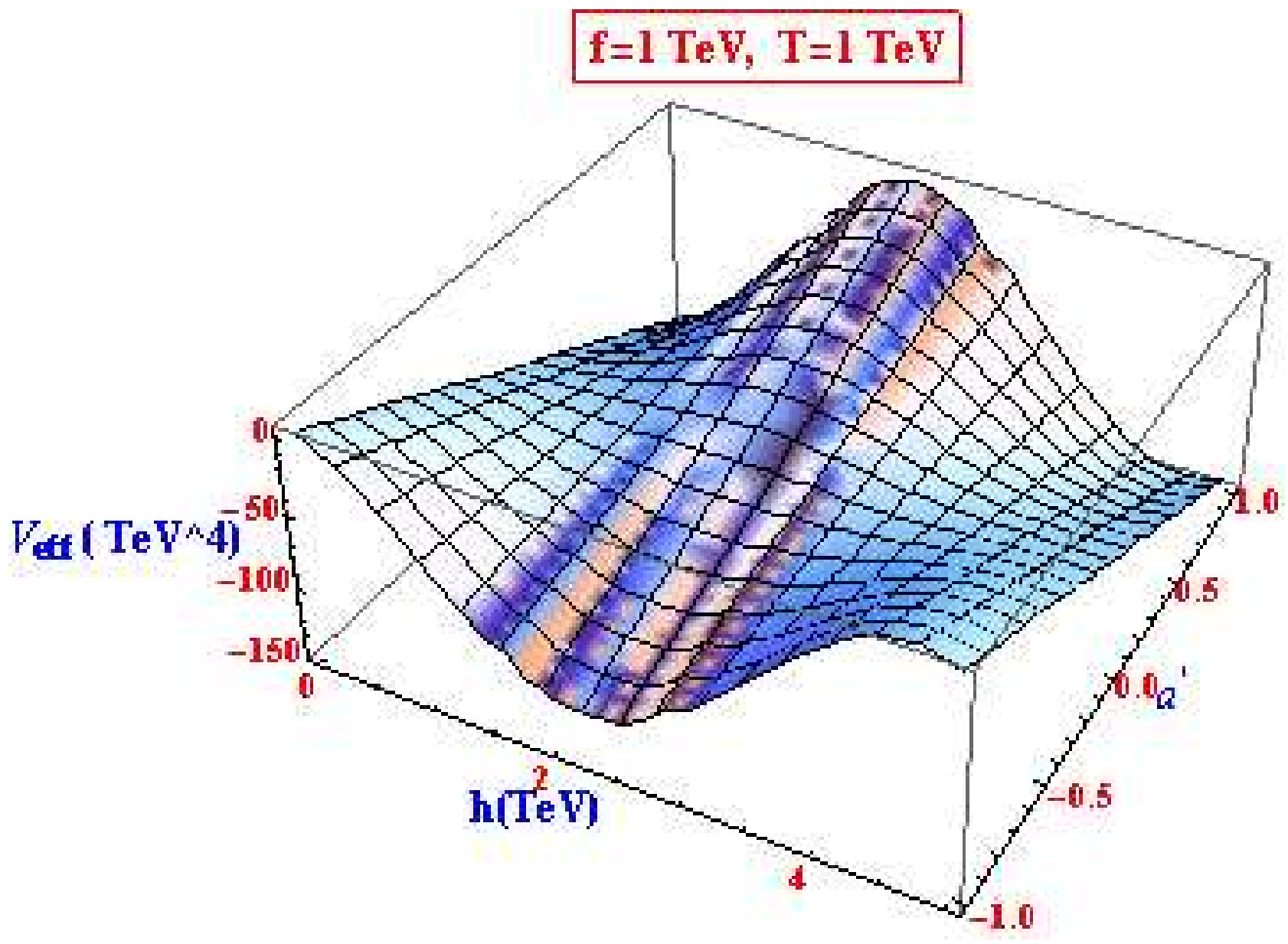

**Fig.5 Variation of the global structure of the one loop order effective potential in the littlest Higgs model without T-parity with the UV completion factor $a'$ keeping the value of the other UV completion factor $a$, fixed at the value -1.**

Next, we study the effects of temperature on the global structure of the effective potential (Fig.4). The finite-temperature integrals are evaluated numerically without high-T approximation. The graphs show no breaking of symmetry with $a = -1$and $a' = -0.004$ at the position of the first maximum and gradual increase in the value of the potential as the temperature is increased. However, the 3-dimentional plots in Figs. 5 and 6 respectively show that symmetry breaking is possible, if one goes to higher values of $a'$ in the negative direction, keeping $a$ fixed at -1 and to higher values of $a$ in the positive direction, keeping $a'$ fixed at -0.004, at a particular temperature.

Next, we show some results of calculations with T-parity in the model. The calculations are done with f=0.5 TeV. At this value of f, a set of values of the UV completion factors which gives the SM electroweak minimum at zero temperature is $a = -1$ and $a' = -0.02$. For the gauge boson sector we have used the parameter set mentioned in section 5. For the fermion sector, we have considered the $t$ and the $T_+$ fields. The $T_-$ field is not taken into account as it is $h$-independent and gives small contribution through the subdominant logarithmic term.

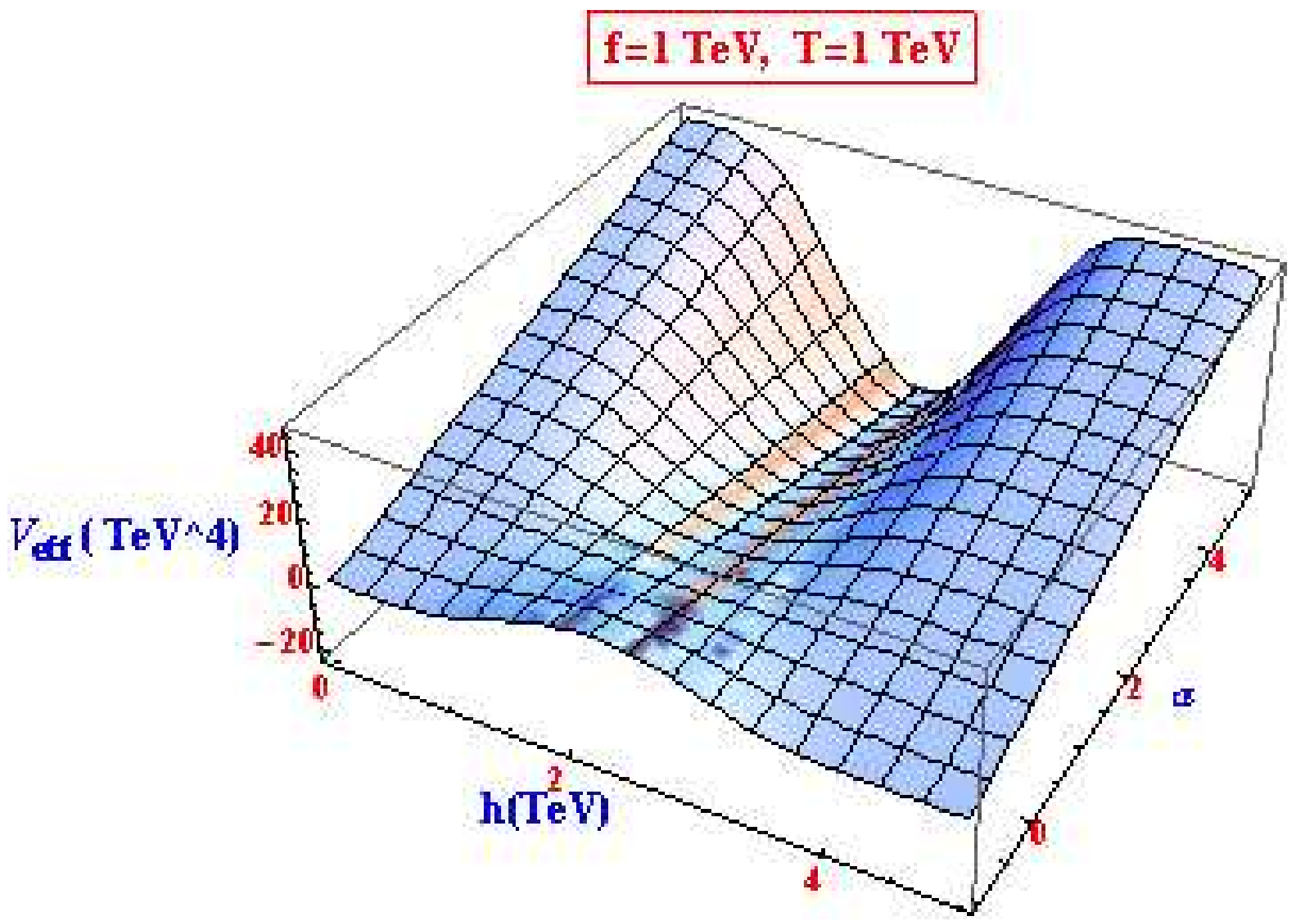

**Fig.6 Variation of the global structure of the one loop order effective potential in the littlest Higgs model without T-parity with the UV completion factor $a$, keeping the value of the other UV completion factor $a'$, fixed at the value -0.004.**

In Fig.7, we show the temperature variation of the effective potential near the SM electroweak minimum. Like in the case of $L^2$HM without T-parity, the SM electroweak symmetry is not restored at high temperature. For this restoration, variation of the values of $a$ and $a'$ are needed, as shown in Figs. 5 and 6.

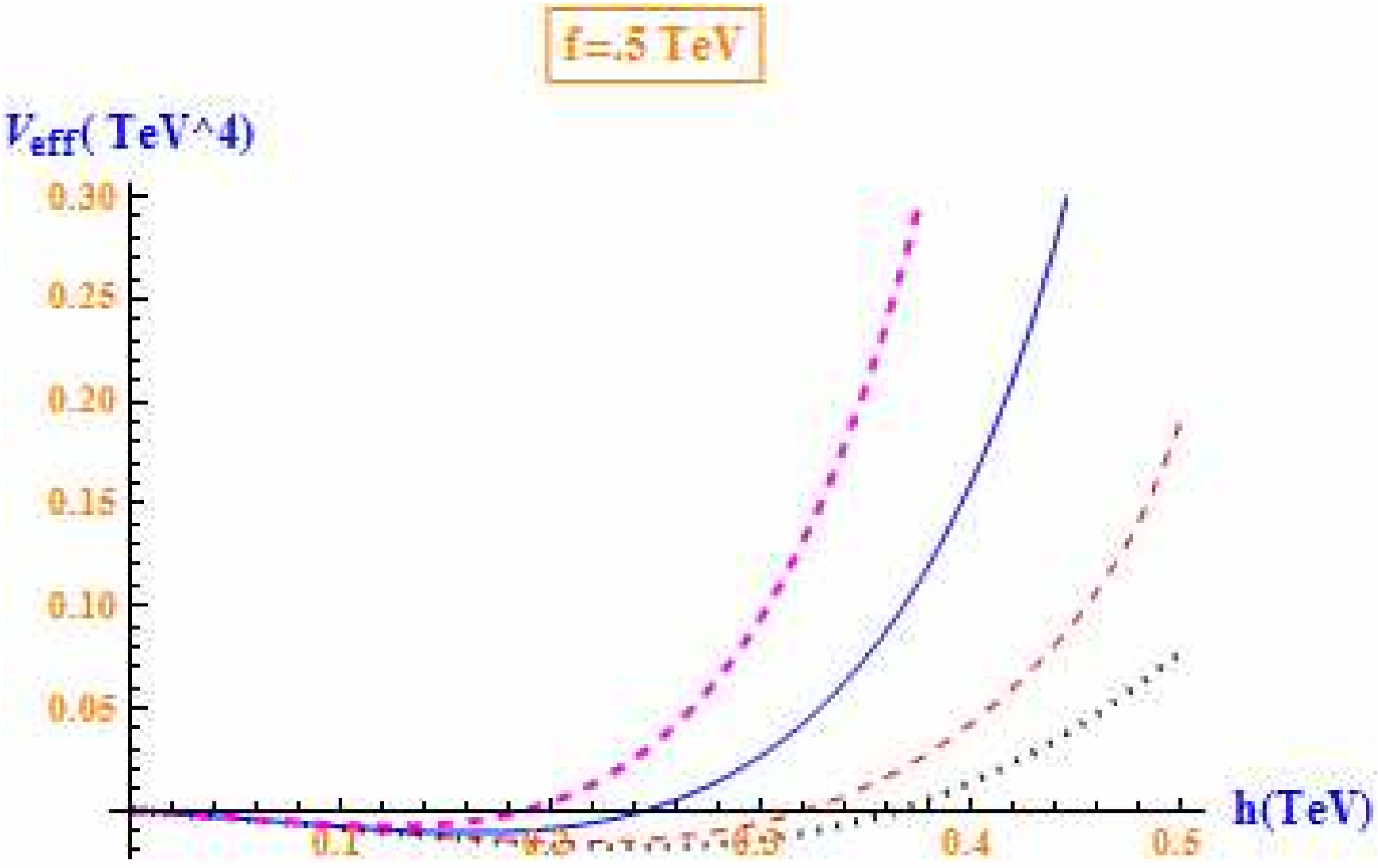

**Fig.7 Effective potential with T-parity near the SM electroweak minimum for various temperatures: T=0 TeV (black-dotted), T=1.0 TeV (red, dashed), T=1.5 TeV (blue, solid), T=2.0(magenta, thick-dashed).**

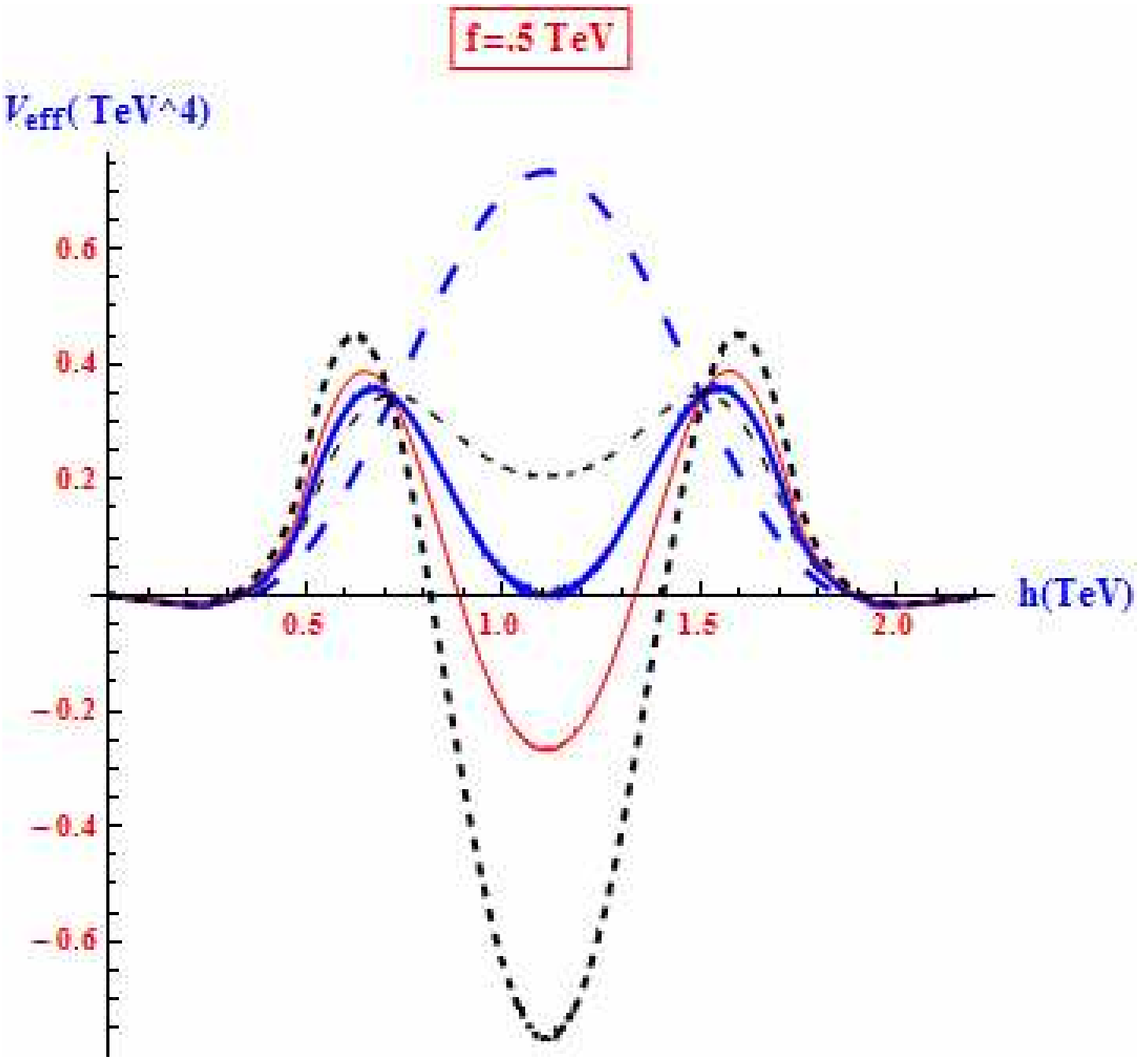

**Fig.8 Finite-temperature effective potential with T-parity at temperatures (from top to bottom): T= 0 TeV (blue, large-dashed), T=0.85 TeV (black, medium-dashed), T= 0.925 TeV (blue, thick-solid), T= 1 TeV (red, solid) and T= 1.1 TeV (black, thick-dashed). The transition temperature, $T_c$ = 0.925 TeV. The values of UV completion factors are here,** $a=-1,\ a'=-0.02.$

We have examined the global structure of finite-temperature effective potential in LHT in Fig.8 and the results here are quite interesting. We find that symmetry breaks as the temperature is increased and the transition temperature is 0.925 TeV. The minimum in the broken phase becomes deeper with the increase in temperature from 1.0 TeV to 1.1 TeV and the minimum for the maximum possible temperature *viz*., T=2.0 TeV, for which we have not plotted the graph, would go too deep to be observed in the figure. The result is similar to the one obtained in Ref. 15 and can be related to the properties of pseudo-Nambu-Goldstone bosons at high temperatures[21]. However, in Ref.15, T-parity was not used in the Little Higgs model, whereas, in our calculation, we have got the result with T-parity. It may be noted that the result obtained in Ref.21 was with a $Z_2$ symmetry, which is there in LHT. The graphs show properties of first-order phase transition, characterised by the existence of false vacua in the symmetric phase, as at temperature 1.0 TeV, shown in Fig. 8.

The physics behind the difference in the behaviour of the effective potential at high temperature in the two cases, *viz*., without T-parity (Fig.4) and with T-parity (Fig.8) can be understood as follows. In our calculation, the UV completion factor for the fermion sector is -0.004 in the former case and -0.02 in the latter case, while for the gauge boson sector it is -1 for both the cases. Consequently, the fermion sector becomes more dominant when there is T-parity than when there is not. The enhanced effective potential in the negative direction in the fermion sector in the case of T-parity makes the thermal mass squared negative at some temperatures, causing symmetry breaking. This is consistent with the earlier observation[6,7] that in Little Higgs models, EWSB is triggered by the large Yukawa coupling between the Higgs and the heavy top quark. We may note that in the case without T-parity also, broken phase may be obtained by artificially increasing the UV factor in the fermion sector in the negative direction, as shown in our three-dimensional plot in Fig.5

The continuous variation of the effective potential as a function of temperature for the case in Fig.8 is shown in Fig.9.

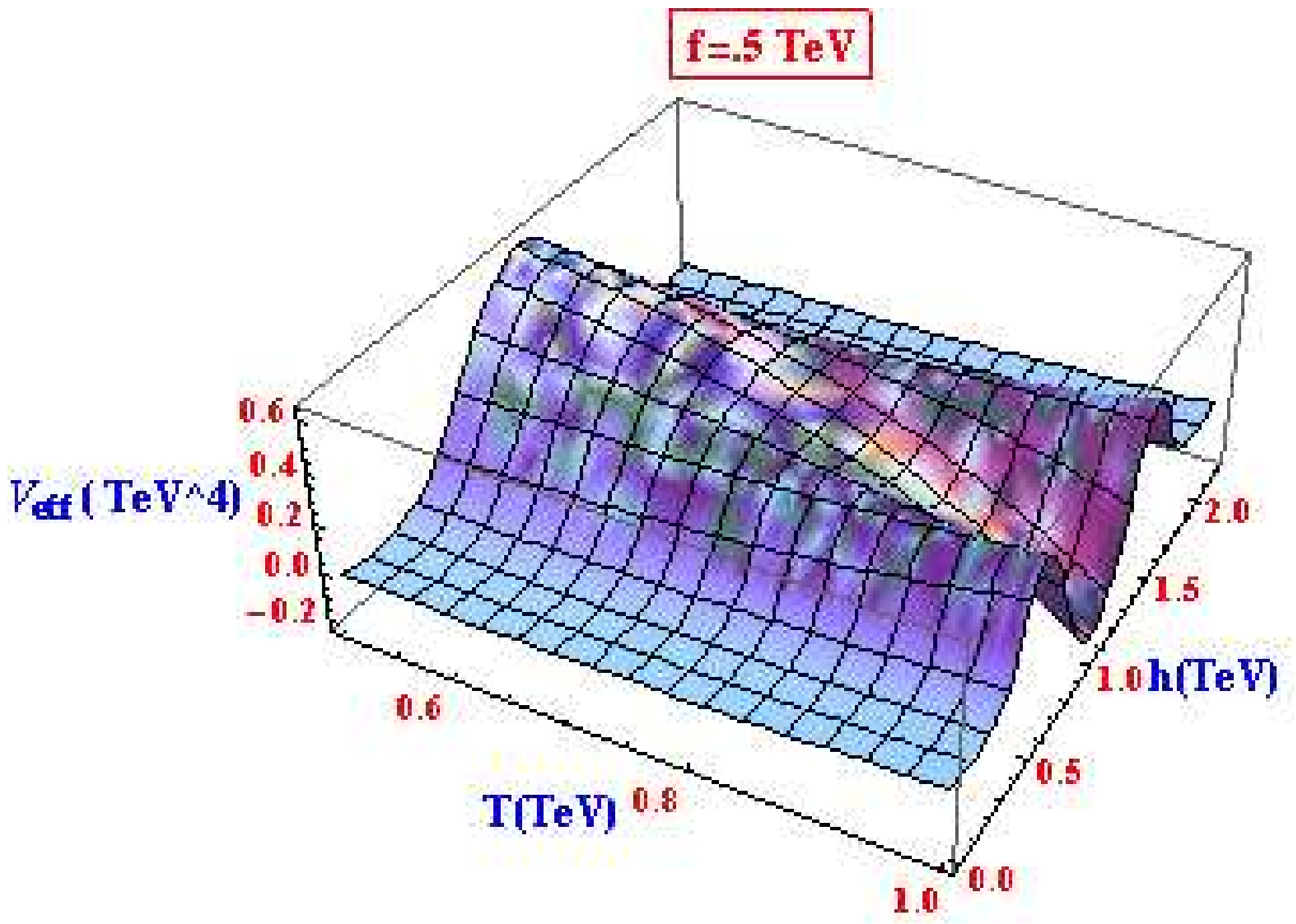

**Fig.9 Variation of the global structure of the one loop order effective potential with temperature in the Littlest Higgs model with T-parity. The values of the UV completion factors are** $a = -1$, $a' = -0.02$, **which gives the Standard Model electroweak minimum at h=0.246 TeV.**

The results given above and obtained by us are based on one-loop order calculation of the effective potential. While full two-loop calculation has not been presented in literature so far, significances of such calculations have been discussed. Inclusion of two-loop calculations has the result of bringing down a large higgs mass(~800 GeV)[12], obtained with $f = 2$, to its current lower bound (~115 GeV). However, we have done our calculations with lower values of $f$, favoured by T-parity. Such small values of $f$ will definitely reduce the importance of the two-loop. We also mention the result of a recent work[22], where it has been observed that in models with strong first-order EWPT, the higgs self-coupling is suppressed relative to its SM value by a factor of 2 or more. This indicates a stronger validity of perturbative calculation in the present model than in the SM, where strong first-order EWPT is not possible for realistic values of the higgs mass.

It is interesting to note the implication of the present observation of the existence of an electroweak broken phase at high temperature on baryogenesis in the early universe. The condition for a strong first order phase transition[14,23], necessary for baryogenesis, is

$$\frac{\varphi_c}{T_c} \gtrsim 1.0, \tag{6.1}$$

where $\varphi_c$ is the higgs vacuum expectation value at the critical temperature $T_c$. Fig.8 shows $\frac{\varphi_c}{T_c} \simeq 1.2$ in our case. Thus a possible non-standard phase transition at high temperature indicated here can favour baryogenesis at $T > T_c$. However, to understand the process of baryogenesis in detail in finite temperature Little Higgs Models one should calculate the sphaleron energy including the heavy gauge bosons. In a recent work[24] on the study of baryogenesis in a broad class of models involving pseudo-Goldstone bosons it has been observed that in such models, the condition (6.1) is consistent with realistic values of the higgs mass, *viz.*, $m_h \geq 115$ GeV.

Also, since we are getting the indication of phase transition in the Littlest Higgs model with T-parity, where the lightest T-odd gauge boson is a possible dark matter candidate, it may be worthwhile to examine the influence of dark matter on baryogenesis in the framework of the present model.

The above comments concern only the doublet higgs sector on which we have focused our attention in the present study, given the fact that in a T-parity symmetric theory, the doublet and triplet higgs sectors do not mix up. However, in the early universe scenario, the triplet higgs sector may have its own influence on phase transition.

## 7. Conclusions

In conclusion, by evaluating the finite-temperature integrals in the one-loop order effective potential in the Littlest Higgs model numerically, without high temperature approximation, we have studied the non-restoration of electroweak symmetry at high temperature, which was already observed in Ref.15 in a different little Higgs model with high temperature approximation. We have found that if the UV completion factors are chosen to be the ones which give the SM electroweak minimum at 246 GeV, then the model with T-parity shows the non-restoration of symmetry and not the one without T-parity. We get indication of strong first order electroweak phase transition in the LHT, with the symmetric phase at a lower temperature than the broken phase. This is a *non-standard* feature of EWPT and it suggests the existence of a phase of broken electroweak symmetry in the early universe at a temperature which is about ten times higher than the symmetry breaking temperature in the Standard Model. It would be nice if these conclusions are further established by LHT with a UV completion theory [25, 26].

**ACKNOWLEDGEMENT**

Two of us (S. A. and B. G.) thank the University Grants Commission, Government of India for granting a major research project (F. No. 32-36/2006(SR)) under which the present work has been done. B. G. acknowledges useful discussions with S. Raichoudhury of Institute of Theoretical Physics, Jamia Milia University, New Delhi.